\documentclass[apl,superscriptaddress,reprint,amssymb,aps,floatfix,showkeys]{revtex4-1}
\usepackage{amsmath}
\usepackage{amsfonts}%
\usepackage{amssymb}%
\usepackage{graphicx}
\usepackage{color}
\usepackage{tabularx}
\usepackage{braket}
\usepackage{epstopdf}

\begin{document}
\title{Axion Detection with Negatively Coupled Cavity Arrays}

\author{Maxim Goryachev}
\email{maxim.goryachev@uwa.edu.au}
\affiliation{ARC Centre of Excellence for Engineered Quantum Systems, University of Western Australia, 35 Stirling Highway, Crawley WA 6009, Australia}

\author{Ben T. McAllister}
\affiliation{ARC Centre of Excellence for Engineered Quantum Systems, University of Western Australia, 35 Stirling Highway, Crawley WA 6009, Australia}

\author{Michael E. Tobar}
\affiliation{ARC Centre of Excellence for Engineered Quantum Systems, University of Western Australia, 35 Stirling Highway, Crawley WA 6009, Australia}

\begin{abstract}

Using eigenmode analysis and full 3D FEM modelling, we demonstrate that a closed cavity built of an array of elementary harmonic oscillators with negative mutual couplings exhibits a dispersion curve with lower order modes corresponding to higher frequencies. 
Such cavity arrays may help to achieve large mode volumes for boosting sensitivity of the axion searches, where the mode volume for the composed array scales proportional to the number of elements, but the frequency remains constant. 
 The negatively coupled cavity array is demonstrated with magnetically coupling coils, where the sign of next-neighbour coupling (controlled with their chirality) sets the dispersion curve properties of the resonator array medium. Furthermore, we show that similar effects can be achieved using only positively coupled cavities of different frequencies assembled in periodic cells. This principle is demonstrated for the multi-post re-entrant system, which can be realised with an array of straight metallic rods organised in chiral structures.

\end{abstract}
\date{\today}
\maketitle

\section{Introduction}

Precision measurements using systems with small dissipation was a pioneering field of research of Vladimir Braginsky\cite{systemssmalldissipation}. As pointed out in his work, such systems are very valuable in many areas of physics since they preserve coherence for very long times and may serve as very sensitive probes of physical quantities, such as material property characterisation\cite{Braginsky1987} and tests of fundamental physics\cite{Braginsky1977}, where the ultimate limit is given by quantum mechanics\cite{Braginsky1992,Braginsky547}. One area gaining considerable recent attention is the development of low-temperature microwave cavities, with high-Q factors and low noise readouts to search for the dark matter axion\cite{Peccei:1977aa,AxionDM}.

Axion searches in the microwave frequency band (greater than $1$GHz) poses certain difficulties related to contradicting requirements. As the frequency space pushes towards higher frequencies, detector cavity sizes shrink decreasing the mode volume and corresponding sensitivity. An obvious solution to the problem is to increase the number of detecting cavities~\cite{Kinion:thesis,Multicav2,Multicav3,PhysRevLett.116.161804,resp1}. Unfortunately, this immediately leads to an increase in system complexity as such a system requires the need for additional amplifiers, microwave lines, etc. So, there is a need for solutions that can lead to an increase in the axion detecting mode volume while preserving the number of detecting systems and high resonant frequency. In particular, high mass axions yielding high frequency photons ($>15$GHz) are motivated theoretically~\cite{SMASH}, and by some curious experimental results~\cite{Beck}, but are as yet largely unprobed, although some recent proposals and experiments have begun to move in this direction~\cite{ORGANPaper,CULTASK,BenBs,MADMAX,ADMXHF,Rybka:2015aa,Barbieri:2017aa}. 

In this work, we propose new methods to scale the mode volume based on negative coupling between cavities. This approach gives an advantage over simultaneous averaging over the same number of independent cavities (giving the same scaling law), as the proposed approach requires only one measurement system.

\section{Axion Electrodynamics}

Axion electrodynamics may be considered as an extension to the classical electromagnetism with an additional Lagrangian term coupling the axion scalar field $a$ and the electric $\mathbf{E}$ and magnetic $\mathbf{B}$ components of the electromagnetic field\cite{Wilczek:1987aa}:
\begin{multline}
\begin{aligned}
	\label{EP001E}
		\mathcal{L} = \kappa a \mathbf{E}\cdot \mathbf{B},
\end{aligned}
\end{multline}
where $\kappa$ is the coupling strength. The conventional approach~\cite{Sikivie:1983aa,SecondSikivie,Asztalos:2010aa,ADMX2} to detect the presence of axions is to apply the strongest available external magnetic field, $\mathbf{B}_0$, and measure photons created by the three particle interaction (\ref{EP001E}), the so-called axion-two photon coupling. The measurement is usually done with a single photonic cavity as a probe antenna, so the interaction reduces to direct coupling between axions and cavity photons:
\begin{multline}
\begin{aligned}
	\label{C005CO}
	\displaystyle  H_i = g_\text{eff} (b+b^\dagger)(c+c^\dagger),
\end{aligned}
\end{multline}
where $c$ ($c^\dagger$) and $b$ ($b^\dagger$) are creation (annihilation) operators for the cavity mode ($[c,c^\dagger]=i$) and axions. The coupling coefficient $g_\text{eff}$ is proportional the  DC field and is a function of cavity geometry and fundamental parameters~\cite{McAllister:2016aa}:
\begin{multline}
\begin{aligned}
	\label{C006CO}
	\displaystyle  g_\text{eff} \sim \omega_c VB_0^2Q_L C_\text{EM},
\end{aligned}
\end{multline}
where $\omega_c$ is the cavity resonance frequency, $V$ is the mode volume of the cavity, $Q_L$ is the loaded quality factor and $C_\text{EM}$ is the unity scaled electromagnetic form factor that is a scaled overlap integral between the external field and the cavity mode. This parameter depends only on the cavity geometry but not on its volume, thus it stays constant when the cavity is scaled. On the other hand, $C_\text{EM}$ depends on the form of a mode used for the detection. For a typical haloscope with a uniform, static magnetic field, $\mathbf{B}_0$, along a single axis, the electric field generated by axion to photon conversion is parallel to this axis. Since, for detection purposes, axions may be thought of as a space uniform field the maximally sensitive mode is the one having the most uniform structure and the lowest number of nodes, i.e. the lowest order mode. Thus, the aim of the detector design for the axion search is to maximise the mode volume $V$ and $C_\text{EM}$ for the given frequency $\omega_c$.

\section{Inductively Coupled Cavity Arrays}

\subsection{Theoretical Principle}

Instead of one isolated cavity, it is possible to consider a set of a regular one dimensional chain of linearly coupled cavities. Each cavity supporting a particular mode may be considered as a harmonic oscillator. A particularly instructive model is an $LC$ circuit, characterised by an ideal capacitance $C$ and inductance $L$. 
The next-neighbouring individual elements of this chain are coupled via mutual inductance $M$. The Hamiltonian of the chain may be written in the form\cite{Lievens:2013aa}:
\begin{multline}
\begin{aligned}
	\label{LP001R}
		\displaystyle H = \omega_0\sum_j c_j^\dagger c_j - g \sum_j (c_j^\dagger c_{j+1} + c_{j+1}^\dagger c_{j}),
\end{aligned}
\end{multline}
where $\omega_0^{-2} = LC$, $g^{-2} = MC$. In this model, we limit coupling to only nearest neighbours, which may be ensured explicitly or is typically the case in most practical realisations {\color{black} where coupling drops either exponentially or inverse polynomially with the distance between elements.}
It is important to underline that the cavity-cavity coupling strength $g$ and thus the mutual inductance $M$ could be both negative or positive depending on the mutual winding of neighbouring inductors. More precisely, the mutual inductance can be designed to be in the region $[-L,L]$.

 It is straightforward to obtain a dispersion relationship from equation (\ref{LP001R}) by substituting the wave solution $\phi_i = A\exp{(-ikj-i\omega t)}$:
\begin{multline}
	\label{LP002R}
\begin{aligned}
		\omega^2 = \frac{1}{{CL}} - \frac{2}{{MC}}\cos k = \omega_0^2\big(1 - 2\eta\cos k\big),
\end{aligned}
\end{multline}
where $\eta = L/M$, $k$ is the wave number and $\omega$ is the angular frequency of a wave propagating along the chain. Here parameter $\eta\in[-1,1]$ due to limits imposed on $M$. If, for example, $\eta = 1/2$, then we obtain a dispersion relationship of a simple vibration lattice of masses and springs where, for small wave numbers, $\omega\sim k$. Although, it is more interesting to consider the parameter subspace where $\eta<0$. In the case when $\eta = -1/2$, $\omega$ spans between $\sqrt{2}\omega_0$ for $k=0$ and $0$ for $k=\pi/2$.

What is important about $\eta<0$ is that the lower wave number waves correspond to higher frequencies and vice a versa. This property can be exploited in an axion search in order to get higher mode volumes at high frequencies, without resorting to moving to higher order modes with decreased form factors. Indeed, the fundamental mode of the chain is the one that has the highest sensitivity to the detectable axion signal, so by reversing the sign of the coupling, one can shift this mode from the lowest to the highest frequency achievable with such a chain. The inverse is true for the highest order mode that has the lowest axion sensitivity. 

Under the considered experimental conditions, {\color{black}to quantify the sensitivity to axions of a certain mode defined on a chain of cavities in terms of eigenvalues of system (\ref{LP001R}), one needs to calculate an overlap between eigenvectors of this mode and sensitivities of single cavities.} The product of overall chain form factor and volume may be viewed as a figure of merit for this, and maybe be expressed for its $k$th order mode in terms of the form factor and volume of each element of the chain $C^\text{(j)}_\text{EM}\times\text{V}^\text{(j)}_\text{element}$ and the value of normalised eigenvectors $v^\text{(j)}_{k}$ of the $k$th order mode:
\begin{multline}
	\label{LP003R}
\begin{aligned}
		\hat{C}^\text{(k)}_\text{EM}\times\text{V}_\text{chain} = \sum_{j=1}^N C^\text{(j)}_\text{EM}\times\text{V}^\text{(j)}_\text{element}~v^\text{(j)}_{k},
\end{aligned}
\end{multline}
where $N$ is the number of cavities in the chain. From this relation it follows that for identical cavities the sensitivity is maximised for the mode with the largest sum of elements of the corresponding eigenvector. This is fulfilled for the uniform mode giving the sum equals to one:
\begin{multline}
	\label{LP004R}
\begin{aligned}
		\hat{C}^\text{(k)}_\text{EM}\times\text{V}_\text{chain} = \text{N}~C_\text{EM}\times\text{V}_\text{element}.
\end{aligned}
\end{multline}
For non uniform modes the divergence from this relation may be expressed as a ratio:
\begin{multline}
	\label{LP005R}
\begin{aligned}
		\xi_k = \frac{\hat{C}^\text{(k)}_\text{EM}\times\text{V}_\text{chain}}{\text{N}C_\text{EM}\times\text{V}_\text{element}}.
\end{aligned}
\end{multline}
Here, $\xi_k$ is a numeric value of $\leq 1$, where the value of unity represents completely uniform modes and is the ideal value, in practice this value needs to be maximised.

\begin{figure}[h!]
     \begin{center}
            \includegraphics[width=0.45\textwidth]{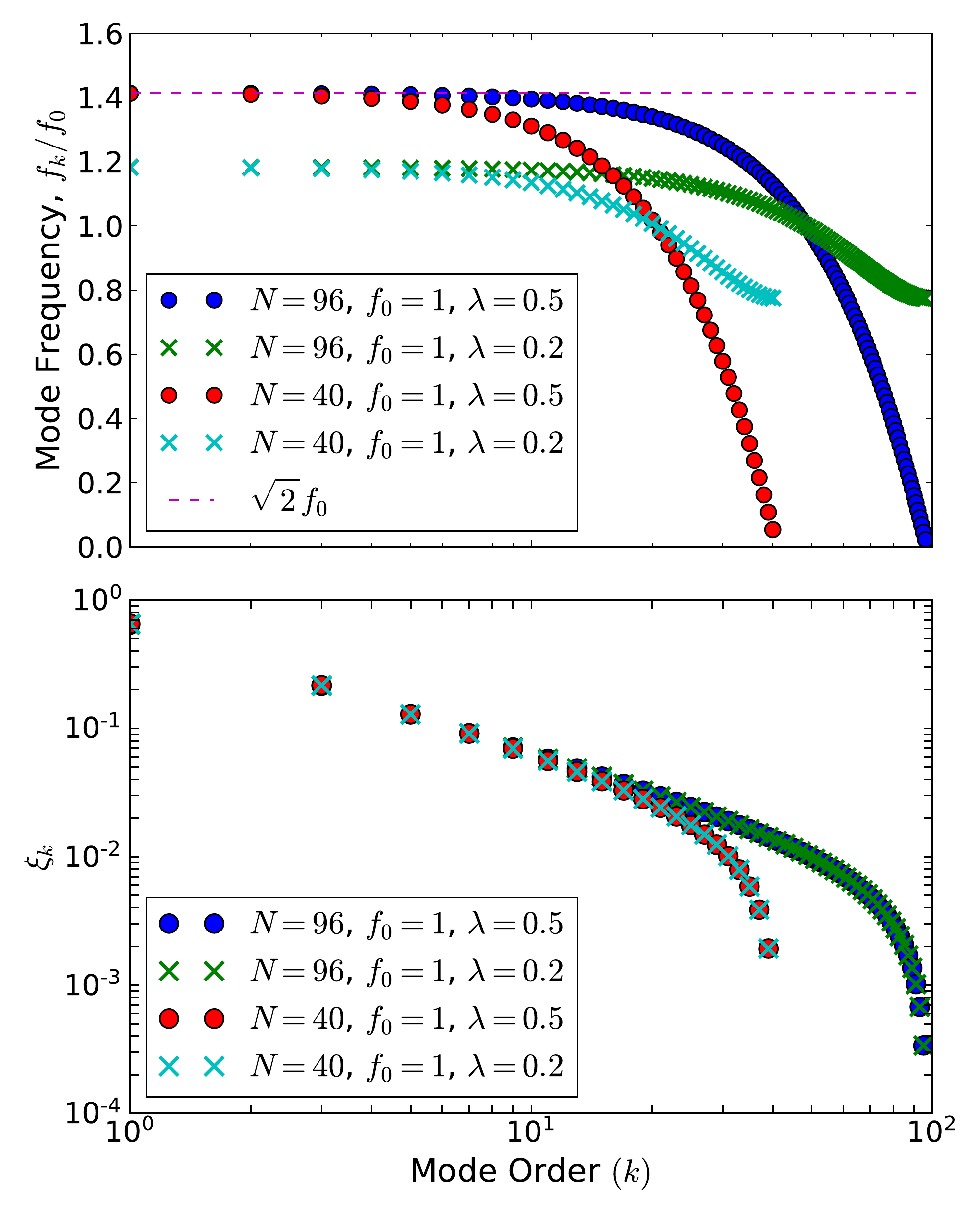}
            \end{center}
    \caption{Resonance frequency (A) and mode divergence factor $\xi_k$ (B) as a function of mode order $k$ for a chain of $N$ cavities.}%
   \label{scale1}
\end{figure}

To demonstrate the effect of negative coupling on resonance frequencies of a chain of cavities, we calculate the eigenfrequencies of such a system. The distribution of resonance frequencies for a system with $N$ cavities and different coupling parameters $\eta$ as a function of mode order $k$ is shown in Fig.~\ref{scale1} (A). The plot illustrates the fact that higher frequency modes correspond to lower order numbers $k$ and, thus, to larger mode volumes that can be observed in Fig.~\ref{scale1} (B). The maximum achievable frequency for the chain is $\sqrt{2}\omega_0$, this follows from the dispersion relationship (\ref{LP002R}). For higher dimensional structures with dimensionality $D$, the frequency scales as $\sqrt{1+D}\omega_0$.

Fig.~\ref{scale1} (B) demonstrates divergence factor $\xi_k$ calculated as a sum over mode eigenvectors. Here, even order modes for all parameter systems give zero parameter values as positive and negative parts of eigenvectors cancel. Also, even the lowest order mode does not reach the limit of $\xi=1$. This result is explained by the fact that due to the open boundary conditions imposed on both ends of the chain, none of these modes are uniform showing decay of intensity from the chain centre to its ends. 
Another important conclusion is that the overall mode volume increases with $N$, thus, it is possible to scale this parameter (and all results that depend on $\xi_n$) with $N$.

\begin{figure}[h!]
     \begin{center}
            \includegraphics[width=0.45\textwidth]{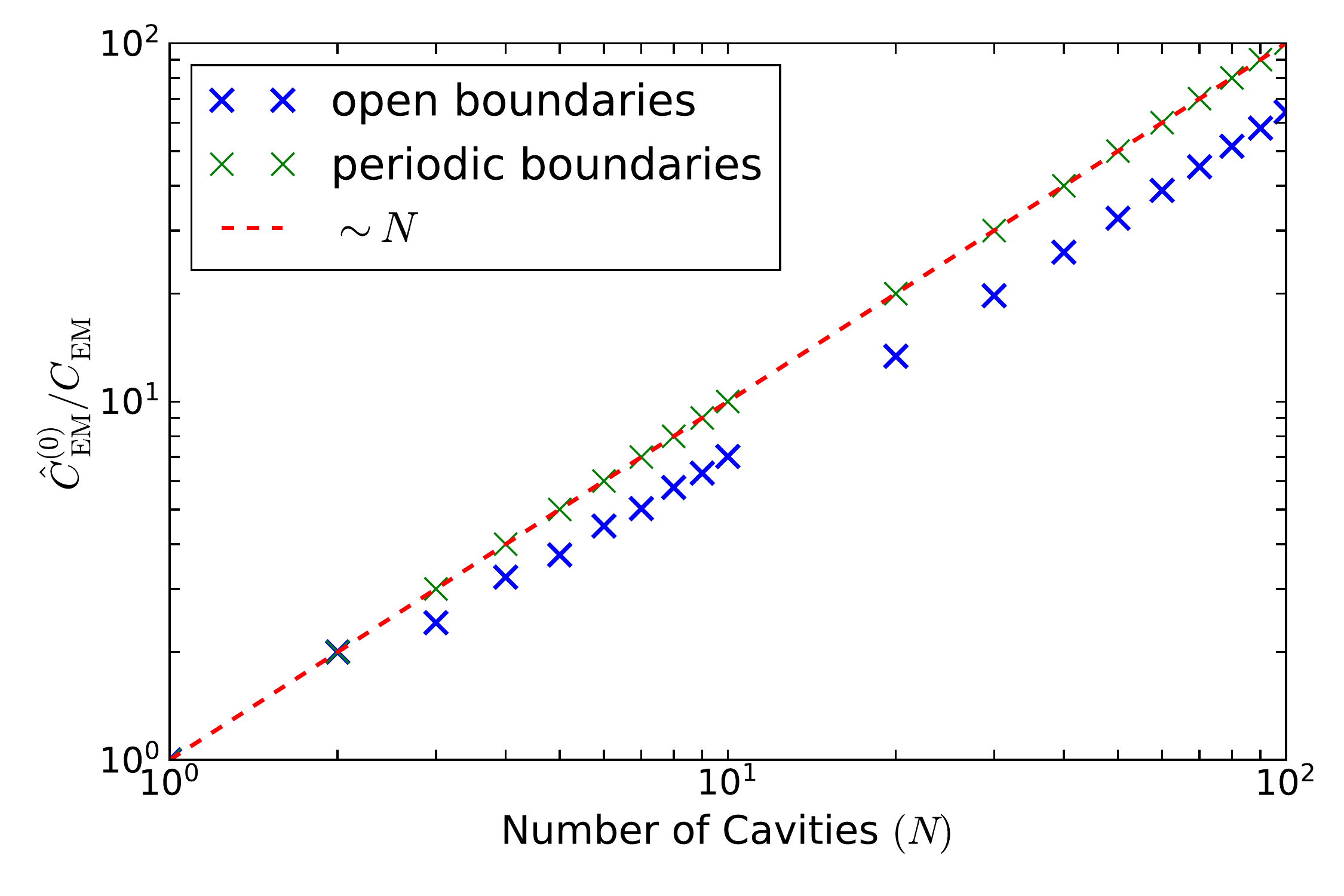}
            \end{center}
    \caption{Form factor of the zero order mode (highest frequency) of a chain of cavities as a function of its length for open and periodic boundaries.}%
   \label{scale2}
\end{figure}

Scaling of the form factor and volume product was calculated for the first mode, $k=0$, with a varying number of cavities, $N$, for two types of boundary conditions. The results are shown in Fig.~\ref{scale2}, which demonstrates the factor of ${N}$ scaling law. The same result can be obtained by power combining $N$ independent cavities, but this introduces many complex technical challenges~\cite{Kinion:thesis,Multicav2,Multicav3}. Recent work suggest that post-processing the data acquired from $N$ independent cavities may yield a small improvement over the system proposed here, although, this would require $N$ independent measurement systems~\cite{XSWISP}. Additionally, the figure shows that the chain with open boundary conditions has suboptimal form factors due to  non-uniformity caused by edge effects. 

A possible complication with the proposed system of a negatively coupled chain of cavities is high density of modes towards the zeroth order. This problem can make it difficult to tune the most sensitive probe mode {\it in situ} due to high mode density. However, in principle, if all cavity frequencies, $\omega_0$, are tuned synchronously, the whole dispersive curve moves up or down with the constant separation between the chain modes. This implies no avoided crossings or other mode interaction effects are present to limit the tunability of the most sensitive mode. Tuning of all cavities simultaneously can be achieved if cavities are made of individual electromagnetically resonant elements enclosed in a shared conducting movable surface. An example of such structure is a multi-element re-entrant type structures described below where one controls post gaps and thus the equivalent capacitance $C$ and thus the $\omega_0$ for all elements. 


\subsection{Physical Implementation}

A system of coupled cavities is often realised using 2D split ring resonators or similar structures making a base for implementation of metamaterials. Recently, a 3D version of this resonator, a multi-post reentant cavity\cite{reen0,reen1,reen2}, has been proposed as a controllable base for metastructures~\cite{Goryachev:2015aa,Goryachev:2015ab,Goryachev:2016aa}. Such a system is made up of a number of conducting posts attached to one cavity wall and closely approaching another with their tips. Every post forms a separate resonant structure with equivalent inductance primarily due to the length and capacitance formed by the gap between the tip and the opposite surface. The additional advantage of such system is its high tunability and possibility to control all post gaps simultaneously. {\color{black} Another important property of re-entrant cavities is possibility to achieve quite high Quality Factors. For example, for superconducting re-entrant cavities, $Q$ factors exceeding $10^8$ have been demonstrated\cite{Bassan2008}.}

\begin{figure}[h!]
     \begin{center}
            \includegraphics[width=0.47\textwidth]{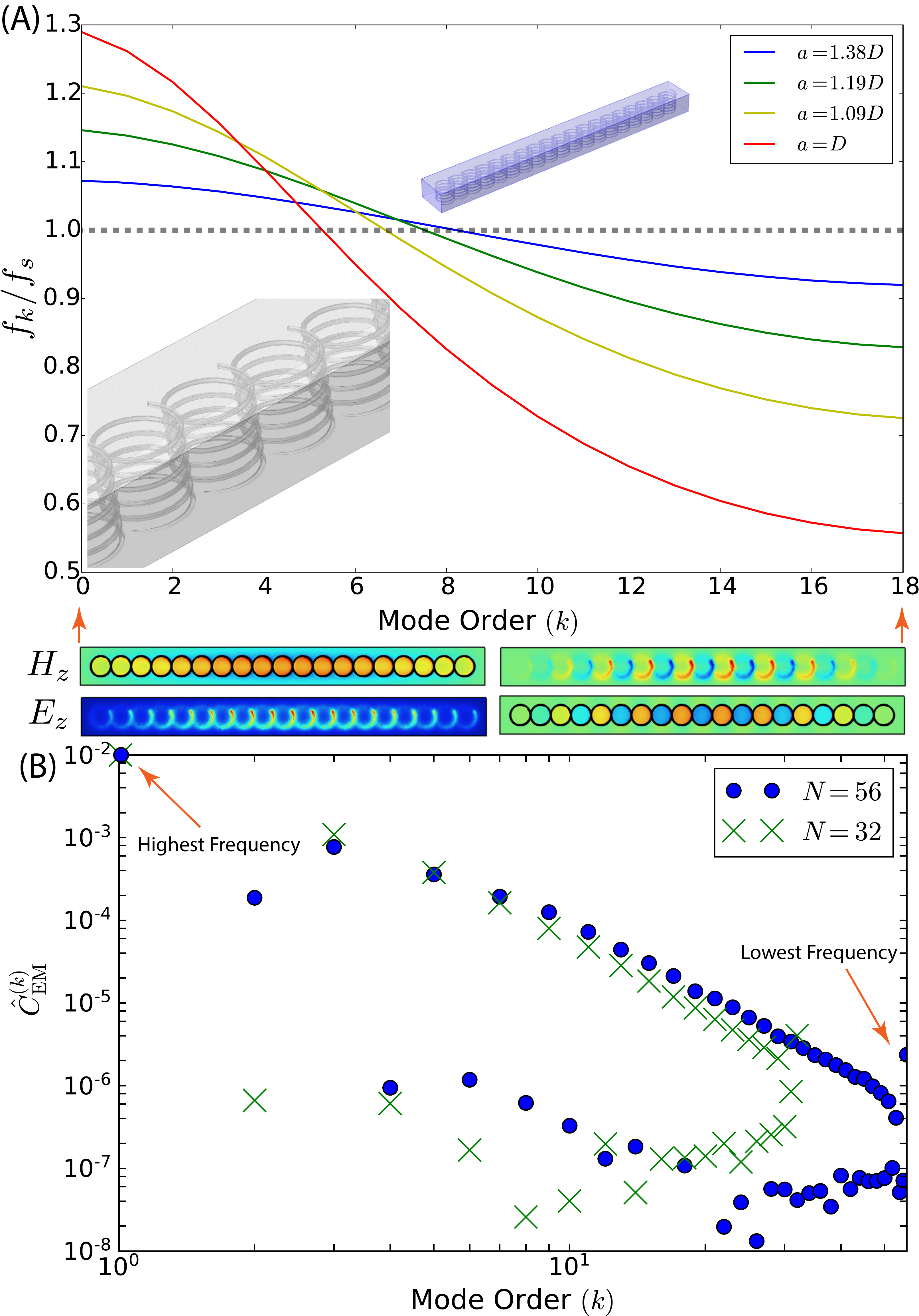}
            \end{center}
    \caption{(A) Mode resonance frequencies as a function of mode order $k$ for different lattice parameters $a$ compared to the coil diameter $D$. Color plots show the electric and magnetic field distribution for the lowest and highest order modes. The insets show 3D views of the system. (B) Filling factors for all re-entrant modes of $N=32$ and $N=48$ chains of coupled coils as a function of mode order $k$.}%
   \label{dispersion}
\end{figure}

It is possible to demonstrate that such a system of parallel straight posts gives negative mutual inductance and, thus, a negative sign of $\eta$. So, a system constructed of a regular grid of such posts effectively composes a right-handed metamaterial with a normal dispersion curve. The sign of coupling between resonators can be reversed by replacing straight posts with coils as shown in the insets of Fig.~\ref{dispersion}. Such a system can be still regarded as a re-entrant structure as the gap between the coils ends and the opposite surface forming capacitance is preserved. Thus, the system inherits one of the most important features of the re-entrant cavity, i.e. high tunability by changing post gap. This feature is important for axion detection as it allows to scan a wide frequency range. Indeed, highly tuneable re-entrant cavities have previously been proposed as a resonator for axion searches~\cite{RCAxion}.

The performance of the coil structure has been confirmed by full 3D Finite Element Modelling (FEM) in COMSOL Multiphysics. The resulting dispersion relationship is shown in Fig.~\ref{dispersion} (A). Four solid curves represent three values of the lattice spacing $a$ relative to the coil outer diameter $D$. Resonance frequencies are scaled to the frequency of a stand alone reentrant coil $f_s$. The color plots demonstrate the field distribution for the highest and lowest order modes. All dispersion curves have a character specific to the negatively coupled cavity structure described above using the harmonic oscillator model. In particular, the highest frequency mode of the chain is characterised by the most uniform distribution of electric and magnetic fields. This fact will result in the strongest axion sensitivity at the highest frequency of the band as seen from calculations of the $0$th mode form factor $\hat{C}^{(k)}_\text{EM}$ in Fig.~\ref{dispersion} (B) demonstrating dependence of the filling factor on the mode order.

It is also observed that the coupling between the chain elements grows with decreasing spacing between the coils. In the limit of large $a$, the dispersion curve is a horizontal line describing the case of uncoupled resonators having the same frequency $\omega_0$. The opposite limiting case is a situation when coils are placed without intermediate gaps, i.e. $a=D$. In fact, the situation with $a<D$ is possible, but it is dominated by other effects that break down the approximation in which coils are considered as separate resonators. Thus, the result cannot be represented by a normal mode decomposition on the simple chain of cavities. Nevertheless, the FEM simulations demonstrate good qualitative agreement between with the negatively coupled next-neighbour model {\color{black}(\ref{LP002R}) shown in Fig.~\ref{scale1} (A). }

\begin{figure}
	\includegraphics[width=\columnwidth]{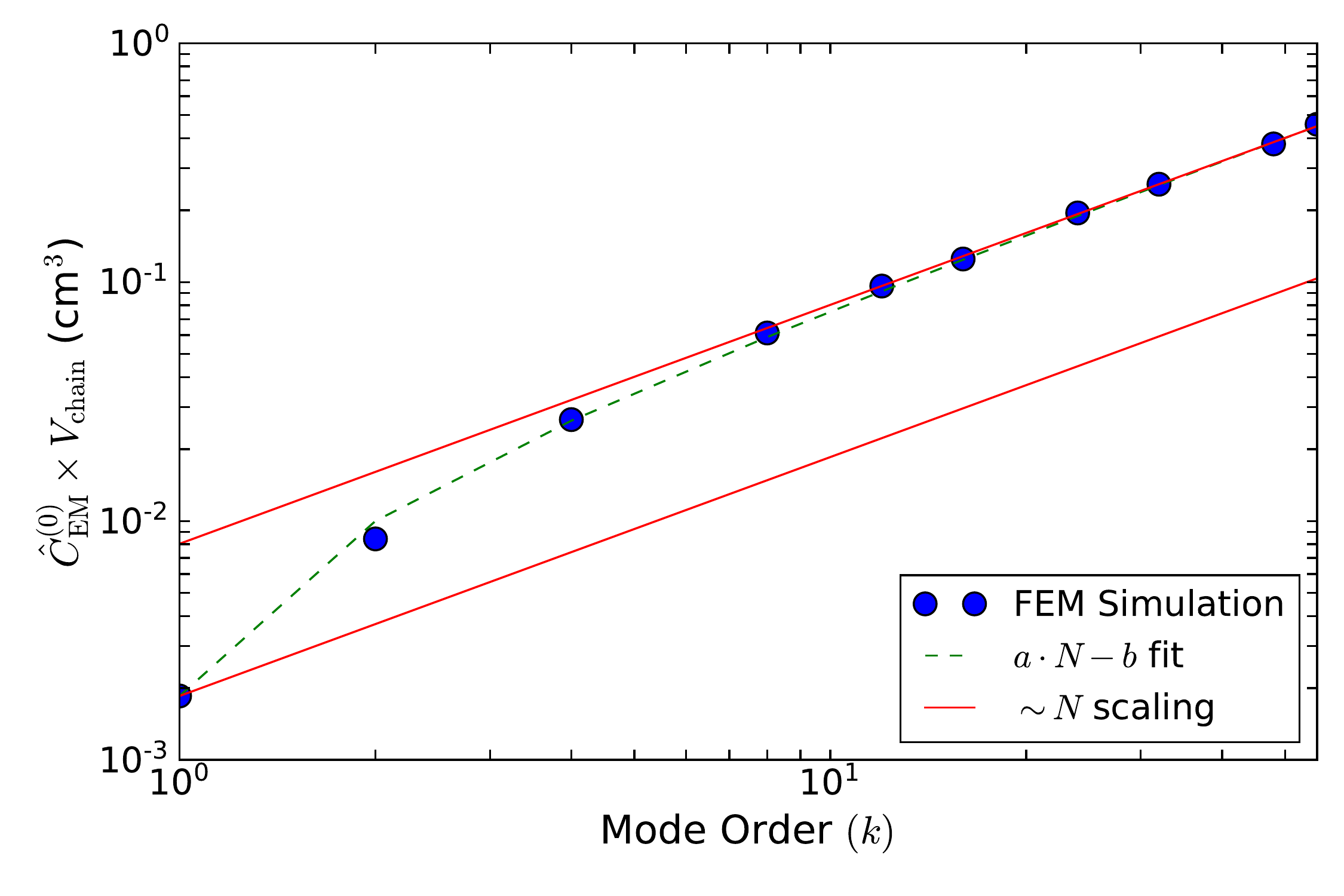}
	\caption{Scaling of the product of the chain form factor and its volume as a function of a number of coils in the chain.}
	\label{fig:CVvN}
\end{figure}

To demonstrate the ability to scale the axion sensitivity, we numerically compute the product of form factor and volume for the most sensitive (zero order) mode, which can be represented as:
\begin{equation}
\displaystyle \text{C}_\text{EM}\times\text{V}_\text{chain}=\frac{\left|\int\text{E}_\text{z}~dV\right|^2}{\int\left|\text{E}_\text{c}\right|^2dV},
\label{eq:CV}
\end{equation}
where $\text{V}_\text{chain}$ is the chain volume and ${E}_\text{c}$ is the total electric field in the cavity {\color{black} and both integrals are taken over the whole system volume}. This quantity characterises sensitivity of the whole structure. The results for different numbers of coils $N$ in the open boundary condition case is shown in Fig.~\ref{fig:CVvN}, which shows that the sensitivity increases proportional to the number of cavities for larger chain lengths  of $N>10$. For the lower values of $N$, the performance is suppressed by a constant value that becomes negligible for $N>10$. This suggests that this suppression is due to edge effects. {\color{black} The axion sensitivity of $N$ independent cavities whose outputs are power combined is represented by the bottom red line. This line demonstrates the $N=1$ case scaled by $N$ as the geometric factor stays constant but the mode volume grows with N, hence so does the overall sensitivity.}

As it is seen from Fig.~\ref{dispersion} (B), the re-entrant mode of the coil with gap exhibits a moderate sensitivity due to significant components of the E field in the $x-y$ plane. Although the proposed array structure improves the performance proportional to $N$, the base sensitivity of a single coil requires some further fine tuning to improve the performance. This may include optimisation of such parameters as the coil radius, thickness, length, axial pitch, etc or the coil structure itself, but such optimisation falls out of the scope of this work. We also note that {\color{black} though the production of many, nominally identical, small coils is well within the limits of standard manufacturing processes, it still will pose certain practical challenges}.

\section{Chiral Cavity Arrays}

The main difficulty related to the physical realisation described in the previous section is manufacturability of the coil-like elements that are used to control the sign of cavity couplings. On the other hand, it is possible to mimic the coil-like (chiral) behaviour between array elements by grouping cavities into quasi-molecules. According to this idea, we consider closely situated inter-coupled groups of cavities as building blocks of sensing arrays and design internal structure of these groups in such a way that desired characteristics of couplings between adjacent groups are achieved. In this case, it is possible to choose the simplest possible structure for the cavity, i.e. simple straight re-entrant post, but make internal group structure more complex in order to mimic negative couplings in the array elements. 
In other words, the complexity of a cavity is transferred onto the complexity of the group structure.

The idea of cavity groups or "quasi-molecules" has been demonstrated in multi-post reentrant cavity systems with straight posts, to realise a photonic topological insulator\cite{Goryachev:2016aa}. This work proposed to implement left-handed and right-handed quasi-molecules with straight (non-chiral) posts by decreasing the gap size under each post when going clock-wise or anti-clockwise across each molecule. With this chiral structure at hand, 1D and 2D arrays of these molecules demonstrated bulk/edge correspondence as well as one-way photon transport. Moreover, the system preserved such nontrivial topological effects when realised on a regular two dimensional grid of posts with accordingly adjusted gaps leading to possible reconfigurability of the overall topological properties. 

\begin{figure}[h!]
     \begin{center}
            \includegraphics[width=0.3\textwidth]{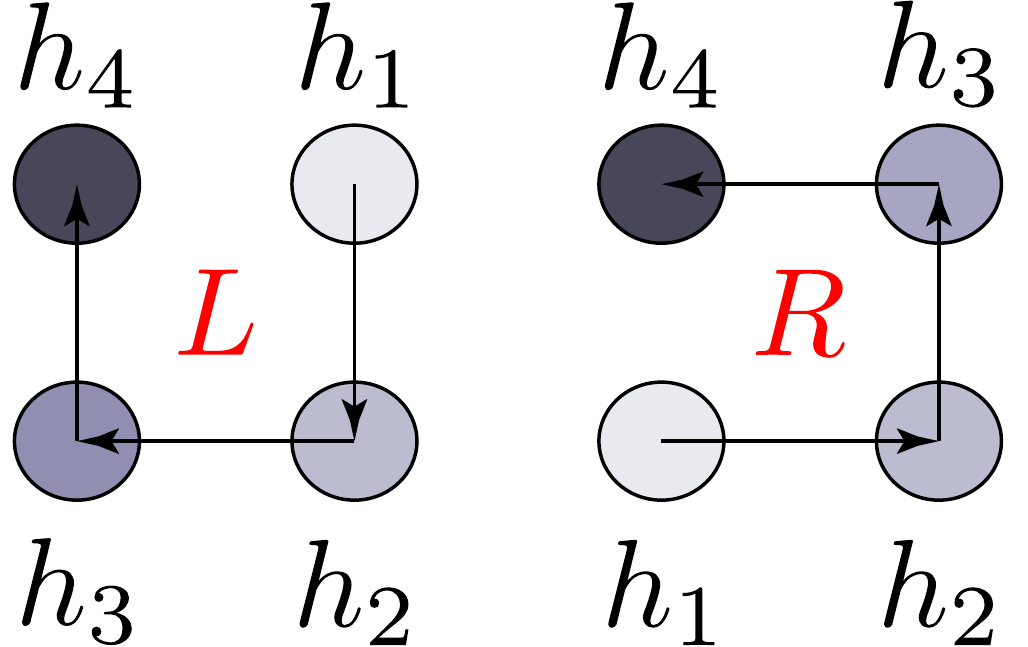}
            \end{center}
    \caption{Two types of reentrant post "molecules" with left or right handedness. Arrows show increase in post gap size $h_i$ and decrease in resonance frequency $\omega_i$ of a corresponding harmonic oscillator.}%
   \label{groups}
\end{figure}

The idea of left handed and right handed groups of straight posts may be applied in the attempt to avoid coil-like posts in the cavity array. Indeed, it is possible to replace one cavity with the coil-like structure with a group of plain cavities having certain sense of handedness and further implementing an array of such groups. Following the first demonstration of this idea\cite{Goryachev:2016aa}, we consider groups of four posts with different gaps $h_i$ distributed clockwise or anti-clockwise (see Fig.~\ref{groups}) implemented on a regular grid. In this section, each post/cavity is modelled as a harmonic oscillator of a certain resonance frequency $\omega_i$ positively coupled to its nearest neighbours. 

\begin{figure}[h!]
     \begin{center}
            \includegraphics[width=0.5\textwidth]{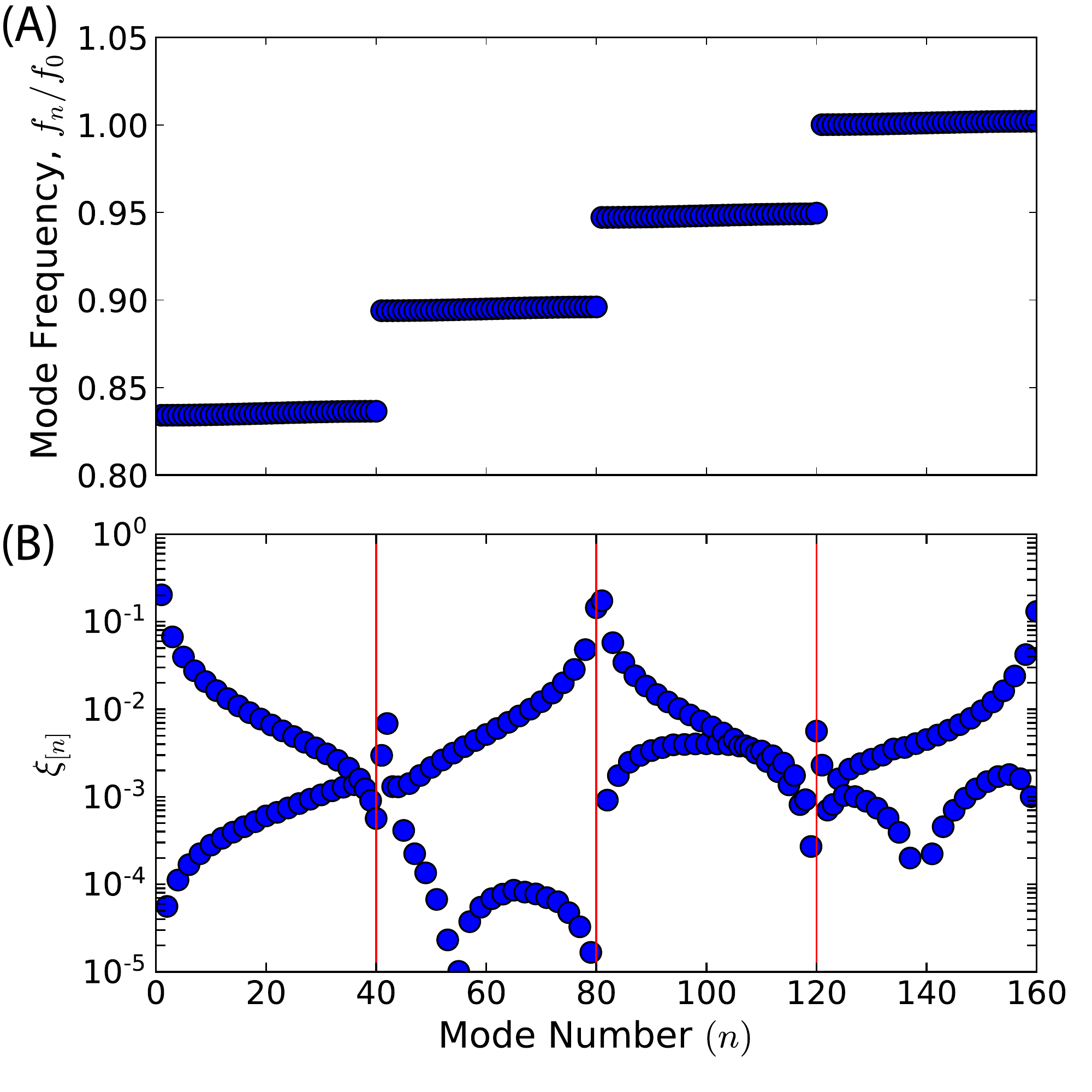}
            \end{center}
    \caption{Scaled resonance frequencies $f_n$ (A) and mode uniformity factors $\xi_n$ (B) for a chain of post groups as a function of mode number $n$. Vertical lines separate the four frequency bands.}%
   \label{divergence}
\end{figure}

For the numerical experiment, we consider a chain of 40 'molecules' with the same handedness and linear change of the resonant frequency of cavities across a single molecule. As the first test, we compute the resonance frequencies $f_n$ and the uniformity of the mode $\xi_n$ (see Eqn. (\ref{LP005R})) for every eigenmode of the array. The described system exhibit multiple frequency bands, for this reason the following discussion will be given in terms of mode number $n$ counted from the lowest to the highest in frequency. This is in contrast to  the mode order $k$ counted from the most uniform to the mode with the maximum number of alternations.
Result are shown in Fig.~\ref{divergence} depicting four frequency bands. In particular, the uniformity, $\xi_{[n]}$, increases towards the highest frequency mode of the second and fourth frequency bands, indicating the desired property of negative couplings between parts of the structure. 

\begin{figure}[h!]
     \begin{center}
            \includegraphics[width=0.5\textwidth]{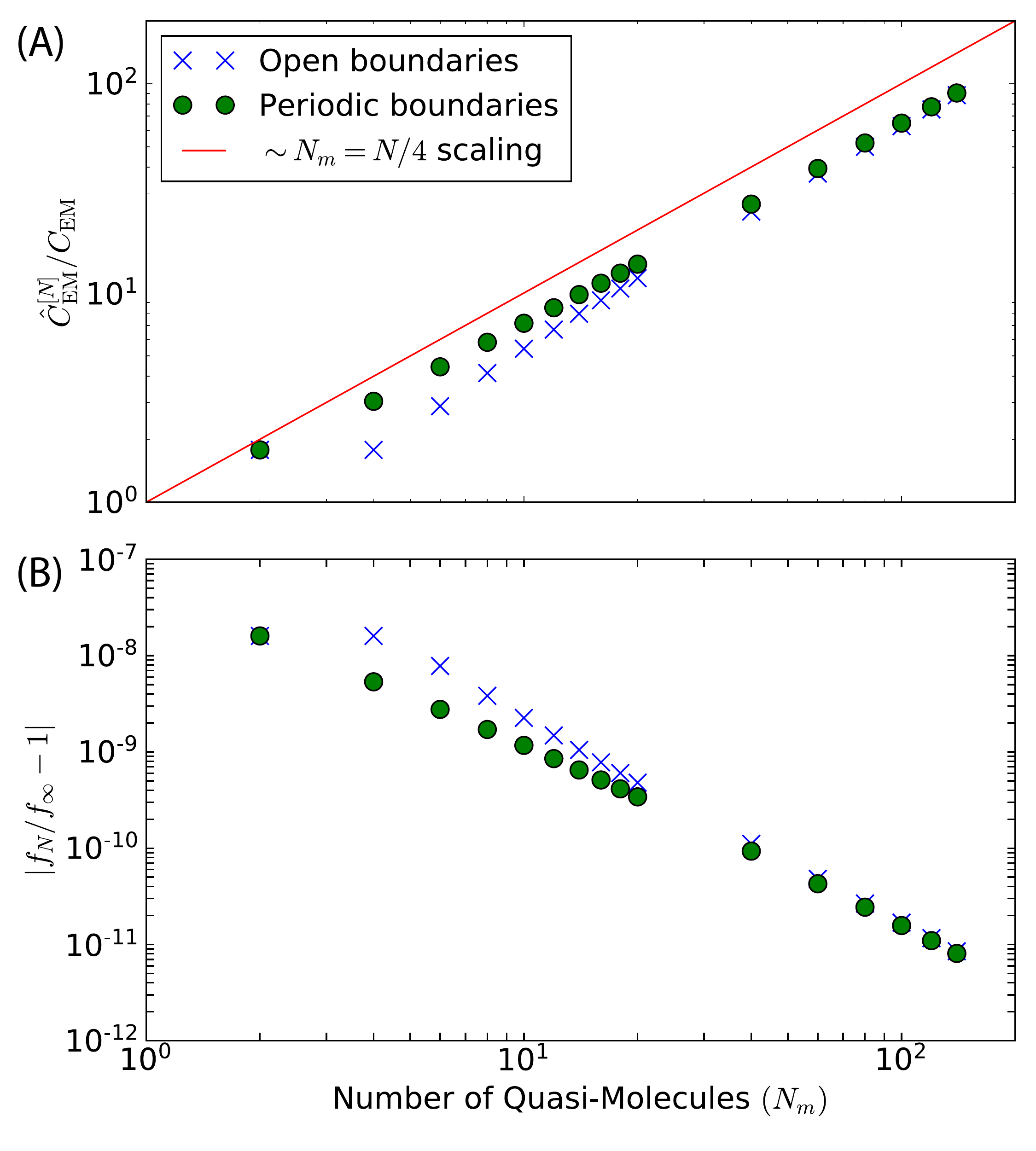}
            \end{center}
    \caption{Form factor (A) and frequency deviation (B) for the highest frequency mode of the chain of cavity molecules as a function of total number of cavity molecules.}%
   \label{scaling}
\end{figure}

In order to investigate the scaling laws for the considered chain, we calculate the form factor of the highest frequency mode as a function of total number of cavity quasi-molecules $N_m$ (here $N_m = N/4$). Fig.~\ref{scaling} demonstrates the result of the form factor scaling for the cases of open and periodic boundary conditions. These conditions are considered with respect to the cavity groups rather than individual cavities. It is possible to optimise the system so that the periodic boundary condition case approaches the $N_m$ line by increasing the difference between resonance frequencies of cavities within a single molecule and adjusting inter-cavity coupling. Finally, a periodic lattice of groups with different handedness may lead to similar effects but its spectrum would be more complex with each of the bands splitting into two.

Fig.~\ref{scaling} demonstrates that the chain of quasi molecules of cavities with the same handedness exhibits a $\sim N_m$ scaling law similar for the case of coil-like elements. However, this solution requires at least four times as many posts or cavities, but does not require a complex coil structure. This may result in higher mode density and more complex mode structures but eliminates the necessity of manufacturing identical coils. 

\begin{figure}[h!]
     \begin{center}
            \includegraphics[width=0.5\textwidth]{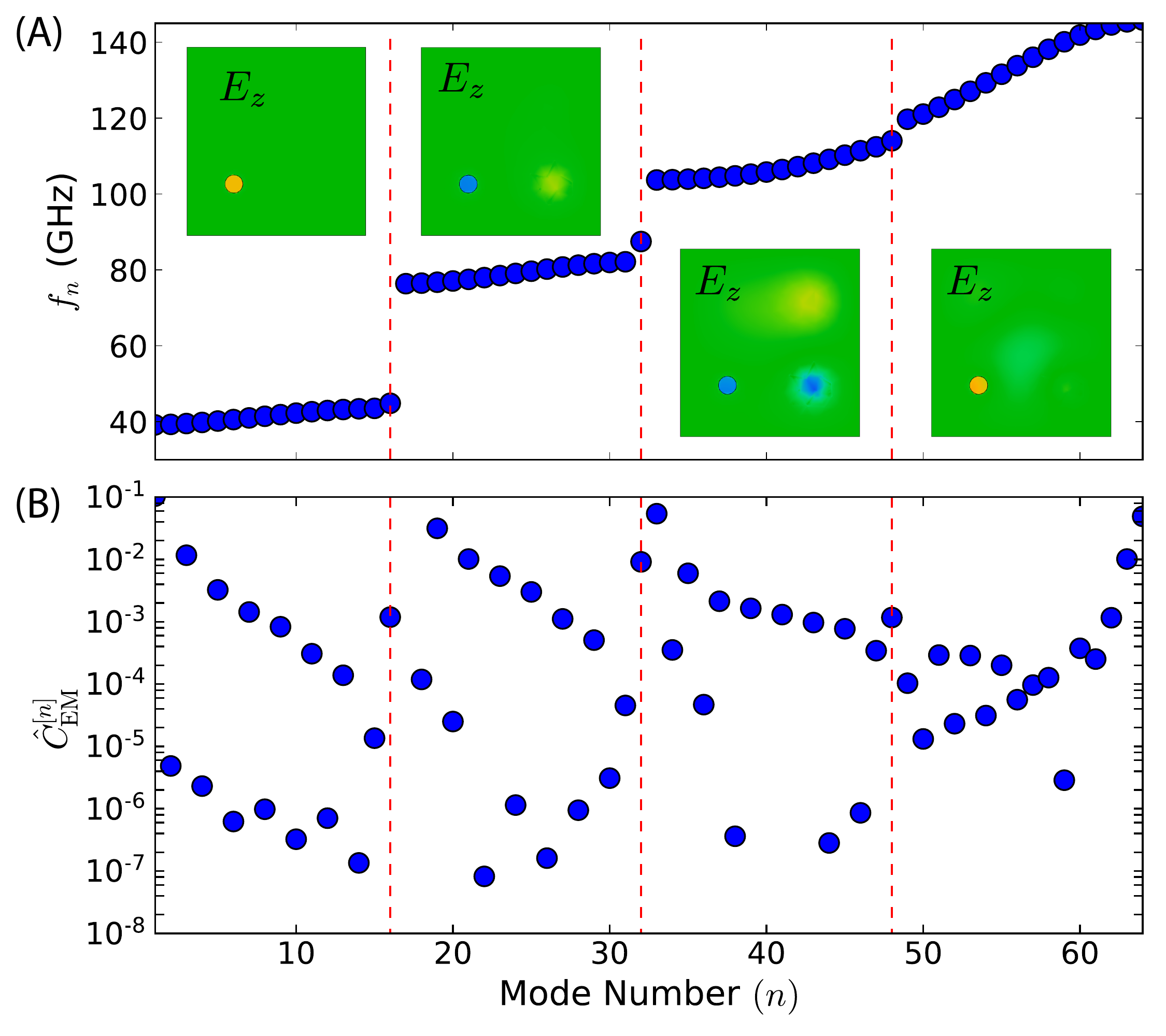}
            \end{center}
    \caption{(A) Resonance frequencies of the FEM modelled chain of post 'molecules'. (B) Corresponding filling factors. The insets show the $E_z$ under posts for one unit cell.}%
   \label{postsFEM}
\end{figure}

To confirm the results of the analysis of the cavity system, we perform full 3D FEM analysis for a straight post realisation of such a system. The system is taken to have 16 quasi-molecules giving 64 as the total number of posts, with the post gaps changed linearly across each unit cell. The simulated resonance frequencies and filling factors are shown in Fig.~\ref{postsFEM} where it is seen that the highest resonance frequency of the fourth band exhibits the largest filling factor, as expected. It should be mention that the system may be further parametrically optimised to achieve higher sensitivity by optimising the post radii, distances and gaps. Such optimisation falls out of the scope of this work.

\section{Conclusion}

In conclusion, we have considered metastructures of negatively coupled cavities that exhibits a dispersion relationship that can help to enhance axion haloscope sensitivity in higher frequency ranges. The structure exhibits the most uniform and hence sensitive modes at the highest frequency of the dispersion curve, higher than the frequency of an uncoupled individual element. Thus, the whole metastructure works as an axion sensitive resonator whose mode volume may be infinitely expanded with the resonant frequency held constant. This technique requires only one measurement system (a set of amplifiers, signal lines, mixers and data acquisition channels), and does not require the synchronization and phase matching of a large array of independent cavities. {\color{black} It also outperforms the power combining method in the large enough number of cavities limit and higher frequency range, as it is not necessary to implement power combiners with arbitrary number of inputs and any additional elements between the cavity and the first amplifier. This is important for high frequencies, $f>20$~GHz as these components significantly drops the efficiency.} Finally, it is numerically shown that this behaviour is possible to achieve in a microwave cavity employing coils as resonant cavity elements or re-entrant posts, grouped into left or right handed unit cells. 
\\
\\
This work was supported by the Australian Research Council Grant No. CE110001013.

\hspace{10pt}

\section*{References}

\begin{thebibliography}{37}%
\makeatletter
\providecommand \@ifxundefined [1]{%
 \@ifx{#1\undefined}
}%
\providecommand \@ifnum [1]{%
 \ifnum #1\expandafter \@firstoftwo
 \else \expandafter \@secondoftwo
 \fi
}%
\providecommand \@ifx [1]{%
 \ifx #1\expandafter \@firstoftwo
 \else \expandafter \@secondoftwo
 \fi
}%
\providecommand \natexlab [1]{#1}%
\providecommand \enquote  [1]{``#1''}%
\providecommand \bibnamefont  [1]{#1}%
\providecommand \bibfnamefont [1]{#1}%
\providecommand \citenamefont [1]{#1}%
\providecommand \href@noop [0]{\@secondoftwo}%
\providecommand \href [0]{\begingroup \@sanitize@url \@href}%
\providecommand \@href[1]{\@@startlink{#1}\@@href}%
\providecommand \@@href[1]{\endgroup#1\@@endlink}%
\providecommand \@sanitize@url [0]{\catcode `\\12\catcode `\$12\catcode
  `\&12\catcode `\#12\catcode `\^12\catcode `\_12\catcode `\%12\relax}%
\providecommand \@@startlink[1]{}%
\providecommand \@@endlink[0]{}%
\providecommand \url  [0]{\begingroup\@sanitize@url \@url }%
\providecommand \@url [1]{\endgroup\@href {#1}{\urlprefix }}%
\providecommand \urlprefix  [0]{URL }%
\providecommand \Eprint [0]{\href }%
\providecommand \doibase [0]{http://dx.doi.org/}%
\providecommand \selectlanguage [0]{\@gobble}%
\providecommand \bibinfo  [0]{\@secondoftwo}%
\providecommand \bibfield  [0]{\@secondoftwo}%
\providecommand \translation [1]{[#1]}%
\providecommand \BibitemOpen [0]{}%
\providecommand \bibitemStop [0]{}%
\providecommand \bibitemNoStop [0]{.\EOS\space}%
\providecommand \EOS [0]{\spacefactor3000\relax}%
\providecommand \BibitemShut  [1]{\csname bibitem#1\endcsname}%
\let\auto@bib@innerbib\@empty
\bibitem [{\citenamefont {Braginsky}\ \emph {et~al.}(1985)\citenamefont
  {Braginsky}, \citenamefont {Mitrofanov}, ,\ and\ \citenamefont
  {Panov}}]{systemssmalldissipation}%
  \BibitemOpen
  \bibfield  {author} {\bibinfo {author} {\bibfnamefont {V.~B.}\ \bibnamefont
  {Braginsky}}, \bibinfo {author} {\bibfnamefont {V.~P.}\ \bibnamefont
  {Mitrofanov}}, , \ and\ \bibinfo {author} {\bibfnamefont {V.~I.}\
  \bibnamefont {Panov}},\ }\href@noop {} {\emph {\bibinfo {title} {Systems with
  Small Dissipation}}}\ (\bibinfo  {publisher} {University of Chicago Press},\
  \bibinfo {address} {Chicago},\ \bibinfo {year} {1985})\BibitemShut {NoStop}%
\bibitem [{\citenamefont {Braginsky}\ \emph {et~al.}(1987)\citenamefont
  {Braginsky}, \citenamefont {Ilchenko},\ and\ \citenamefont
  {BAGDASSAROV}}]{Braginsky1987}%
  \BibitemOpen
  \bibfield  {author} {\bibinfo {author} {\bibfnamefont {V.}~\bibnamefont
  {Braginsky}}, \bibinfo {author} {\bibfnamefont {V.}~\bibnamefont {Ilchenko}},
  \ and\ \bibinfo {author} {\bibfnamefont {K.}~\bibnamefont {BAGDASSAROV}},\
  }\href@noop {} {\bibfield  {journal} {\bibinfo  {journal} {Physics Letters
  A}\ }\textbf {\bibinfo {volume} {120}},\ \bibinfo {pages} {300} (\bibinfo
  {year} {1987})}\BibitemShut {NoStop}%
\bibitem [{\citenamefont {Braginsky}\ \emph {et~al.}(1977)\citenamefont
  {Braginsky}, \citenamefont {Caves},\ and\ \citenamefont
  {Thorne}}]{Braginsky1977}%
  \BibitemOpen
  \bibfield  {author} {\bibinfo {author} {\bibfnamefont {V.~B.}\ \bibnamefont
  {Braginsky}}, \bibinfo {author} {\bibfnamefont {C.~M.}\ \bibnamefont
  {Caves}}, \ and\ \bibinfo {author} {\bibfnamefont {K.~S.}\ \bibnamefont
  {Thorne}},\ }\href@noop {} {\bibfield  {journal} {\bibinfo  {journal} {Phys.
  Rev. D}\ }\textbf {\bibinfo {volume} {15}},\ \bibinfo {pages} {2047}
  (\bibinfo {year} {1977})}\BibitemShut {NoStop}%
\bibitem [{\citenamefont {Braginksy}\ and\ \citenamefont
  {Khalili}(1992)}]{Braginsky1992}%
  \BibitemOpen
  \bibfield  {author} {\bibinfo {author} {\bibfnamefont {V.~B.}\ \bibnamefont
  {Braginksy}}\ and\ \bibinfo {author} {\bibfnamefont {F.~Y.}\ \bibnamefont
  {Khalili}},\ }\href@noop {} {\emph {\bibinfo {title} {Quantum Measurement}}}\
  (\bibinfo  {publisher} {Cambridge University Press},\ \bibinfo {year}
  {1992})\BibitemShut {NoStop}%
\bibitem [{\citenamefont {Braginsky}\ \emph {et~al.}(1980)\citenamefont
  {Braginsky}, \citenamefont {Vorontsov},\ and\ \citenamefont
  {Thorne}}]{Braginsky547}%
  \BibitemOpen
  \bibfield  {author} {\bibinfo {author} {\bibfnamefont {V.~B.}\ \bibnamefont
  {Braginsky}}, \bibinfo {author} {\bibfnamefont {Y.~I.}\ \bibnamefont
  {Vorontsov}}, \ and\ \bibinfo {author} {\bibfnamefont {K.~S.}\ \bibnamefont
  {Thorne}},\ }\href {\doibase 10.1126/science.209.4456.547} {\bibfield
  {journal} {\bibinfo  {journal} {Science}\ }\textbf {\bibinfo {volume}
  {209}},\ \bibinfo {pages} {547} (\bibinfo {year} {1980})}\BibitemShut
  {NoStop}%
\bibitem [{\citenamefont {Peccei}\ and\ \citenamefont
  {Quinn}(1977)}]{Peccei:1977aa}%
  \BibitemOpen
  \bibfield  {author} {\bibinfo {author} {\bibfnamefont {R.~D.}\ \bibnamefont
  {Peccei}}\ and\ \bibinfo {author} {\bibfnamefont {H.~R.}\ \bibnamefont
  {Quinn}},\ }\href {http://link.aps.org/doi/10.1103/PhysRevLett.38.1440}
  {\bibfield  {journal} {\bibinfo  {journal} {Physical Review Letters}\
  }\textbf {\bibinfo {volume} {38}},\ \bibinfo {pages} {1440} (\bibinfo {year}
  {1977})}\BibitemShut {NoStop}%
\bibitem [{\citenamefont {Ipser}\ and\ \citenamefont
  {Sikivie}(1983)}]{AxionDM}%
  \BibitemOpen
  \bibfield  {author} {\bibinfo {author} {\bibfnamefont {J.}~\bibnamefont
  {Ipser}}\ and\ \bibinfo {author} {\bibfnamefont {P.}~\bibnamefont
  {Sikivie}},\ }\href {\doibase 10.1103/PhysRevLett.50.925} {\bibfield
  {journal} {\bibinfo  {journal} {Phys. Rev. Lett.}\ }\textbf {\bibinfo
  {volume} {50}},\ \bibinfo {pages} {925} (\bibinfo {year} {1983})}\BibitemShut
  {NoStop}%
\bibitem [{\citenamefont {Kinion}(2001)}]{Kinion:thesis}%
  \BibitemOpen
  \bibfield  {author} {\bibinfo {author} {\bibfnamefont {D.~S.}\ \bibnamefont
  {Kinion}},\ }\emph {\bibinfo {title} {{First results from a multiple
  microwave cavity search for dark matter axions}}},\ \href
  {http://wwwlib.umi.com/dissertations/fullcit?p3019020} {Ph.D. thesis},\
  \bibinfo  {school} {UC, Davis} (\bibinfo {year} {2001})\BibitemShut {NoStop}%
\bibitem [{\citenamefont {Shokair}\ \emph {et~al.}(2014)\citenamefont {Shokair}
  \emph {et~al.}}]{Multicav2}%
  \BibitemOpen
  \bibfield  {author} {\bibinfo {author} {\bibfnamefont {T.~M.}\ \bibnamefont
  {Shokair}} \emph {et~al.},\ }\href {\doibase 10.1142/S0217751X14430040}
  {\bibfield  {journal} {\bibinfo  {journal} {Int. J. Mod. Phys.}\ }\textbf
  {\bibinfo {volume} {A29}},\ \bibinfo {pages} {1443004} (\bibinfo {year}
  {2014})},\ \Eprint {http://arxiv.org/abs/1405.3685} {arXiv:1405.3685
  [physics.ins-det]} \BibitemShut {NoStop}%
\bibitem [{\citenamefont {{Asztalos}}\ \emph {et~al.}(2000)\citenamefont
  {{Asztalos}}, \citenamefont {{Rosenberg}}, \citenamefont {{Peng}},
  \citenamefont {{Daw}}, \citenamefont {{Kinion}}, \citenamefont {{Hagmann}},
  \citenamefont {{Stoeffll}}, \citenamefont {{van Bibber}}, \citenamefont
  {{Sikivie}}, \citenamefont {{Sullivan}}, \citenamefont {{Tanner}},
  \citenamefont {{Moltz}},\ and\ \citenamefont {{Turner}}}]{Multicav3}%
  \BibitemOpen
  \bibfield  {author} {\bibinfo {author} {\bibfnamefont {S.}~\bibnamefont
  {{Asztalos}}}, \bibinfo {author} {\bibfnamefont {L.}~\bibnamefont
  {{Rosenberg}}}, \bibinfo {author} {\bibfnamefont {H.}~\bibnamefont {{Peng}}},
  \bibinfo {author} {\bibfnamefont {E.}~\bibnamefont {{Daw}}}, \bibinfo
  {author} {\bibfnamefont {D.}~\bibnamefont {{Kinion}}}, \bibinfo {author}
  {\bibfnamefont {C.}~\bibnamefont {{Hagmann}}}, \bibinfo {author}
  {\bibfnamefont {W.}~\bibnamefont {{Stoeffll}}}, \bibinfo {author}
  {\bibfnamefont {K.}~\bibnamefont {{van Bibber}}}, \bibinfo {author}
  {\bibfnamefont {P.}~\bibnamefont {{Sikivie}}}, \bibinfo {author}
  {\bibfnamefont {N.}~\bibnamefont {{Sullivan}}}, \bibinfo {author}
  {\bibfnamefont {D.}~\bibnamefont {{Tanner}}}, \bibinfo {author}
  {\bibfnamefont {D.}~\bibnamefont {{Moltz}}}, \ and\ \bibinfo {author}
  {\bibfnamefont {M.}~\bibnamefont {{Turner}}},\ }in\ \href@noop {} {\emph
  {\bibinfo {booktitle} {APS Meeting Abstracts}}}\ (\bibinfo {year}
  {2000})\BibitemShut {NoStop}%
\bibitem [{\citenamefont {McAllister}\ \emph
  {et~al.}(2016{\natexlab{a}})\citenamefont {McAllister}, \citenamefont
  {Parker},\ and\ \citenamefont {Tobar}}]{PhysRevLett.116.161804}%
  \BibitemOpen
  \bibfield  {author} {\bibinfo {author} {\bibfnamefont {B.~T.}\ \bibnamefont
  {McAllister}}, \bibinfo {author} {\bibfnamefont {S.~R.}\ \bibnamefont
  {Parker}}, \ and\ \bibinfo {author} {\bibfnamefont {M.~E.}\ \bibnamefont
  {Tobar}},\ }\href {\doibase 10.1103/PhysRevLett.116.161804} {\bibfield
  {journal} {\bibinfo  {journal} {Phys. Rev. Lett.}\ }\textbf {\bibinfo
  {volume} {116}},\ \bibinfo {pages} {161804} (\bibinfo {year}
  {2016}{\natexlab{a}})}\BibitemShut {NoStop}%
\bibitem [{\citenamefont {Lee}\ \emph {et~al.}()\citenamefont {Lee},
  \citenamefont {S.W.Youn},\ and\ \citenamefont {Semertzidis}}]{resp1}%
  \BibitemOpen
  \bibfield  {author} {\bibinfo {author} {\bibfnamefont {S.}~\bibnamefont
  {Lee}}, \bibinfo {author} {\bibnamefont {S.W.Youn}}, \ and\ \bibinfo {author}
  {\bibfnamefont {Y.}~\bibnamefont {Semertzidis}},\ }\href@noop {} {\emph
  {\bibinfo {title} {Comment on "Axion Dark Matter Coupling to Resonant Photons
  via Magnetic Field"}}},\ \bibinfo {address} {arXiv:1606.09504}\BibitemShut
  {NoStop}%
\bibitem [{\citenamefont {Ballesteros}\ \emph {et~al.}(2017)\citenamefont
  {Ballesteros}, \citenamefont {Redondo}, \citenamefont {Ringwald},\ and\
  \citenamefont {Tamarit}}]{SMASH}%
  \BibitemOpen
  \bibfield  {author} {\bibinfo {author} {\bibfnamefont {G.}~\bibnamefont
  {Ballesteros}}, \bibinfo {author} {\bibfnamefont {J.}~\bibnamefont
  {Redondo}}, \bibinfo {author} {\bibfnamefont {A.}~\bibnamefont {Ringwald}}, \
  and\ \bibinfo {author} {\bibfnamefont {C.}~\bibnamefont {Tamarit}},\ }\href
  {\doibase 10.1103/PhysRevLett.118.071802} {\bibfield  {journal} {\bibinfo
  {journal} {Phys. Rev. Lett.}\ }\textbf {\bibinfo {volume} {118}},\ \bibinfo
  {pages} {071802} (\bibinfo {year} {2017})}\BibitemShut {NoStop}%
\bibitem [{\citenamefont {Beck}(2013)}]{Beck}%
  \BibitemOpen
  \bibfield  {author} {\bibinfo {author} {\bibfnamefont {C.}~\bibnamefont
  {Beck}},\ }\href {\doibase 10.1103/PhysRevLett.111.231801} {\bibfield
  {journal} {\bibinfo  {journal} {Phys. Rev. Lett.}\ }\textbf {\bibinfo
  {volume} {111}},\ \bibinfo {pages} {231801} (\bibinfo {year}
  {2013})}\BibitemShut {NoStop}%
\bibitem [{\citenamefont {McAllister}\ \emph
  {et~al.}(2016{\natexlab{b}})\citenamefont {McAllister}, \citenamefont
  {Parker}, \citenamefont {Ivanov},\ and\ \citenamefont {Tobar}}]{ORGANPaper}%
  \BibitemOpen
  \bibfield  {author} {\bibinfo {author} {\bibfnamefont {B.~T.}\ \bibnamefont
  {McAllister}}, \bibinfo {author} {\bibfnamefont {S.~R.}\ \bibnamefont
  {Parker}}, \bibinfo {author} {\bibfnamefont {E.~N.}\ \bibnamefont {Ivanov}},
  \ and\ \bibinfo {author} {\bibfnamefont {M.~E.}\ \bibnamefont {Tobar}},\ }in\
  \href {https://inspirehep.net/record/1500189/files/arXiv:1611.08082.pdf}
  {\emph {\bibinfo {booktitle} {{12th Patras Workshop on Axions, WIMPs and
  WISPs (AXION-WIMP 2016) Jeju Island, South Korea, June 20-24, 2016}}}}\
  (\bibinfo {year} {2016})\ \Eprint {http://arxiv.org/abs/1611.08082}
  {arXiv:1611.08082 [hep-ex]} \BibitemShut {NoStop}%
\bibitem [{\citenamefont {Chung}(2016)}]{CULTASK}%
  \BibitemOpen
  \bibfield  {author} {\bibinfo {author} {\bibfnamefont {W.}~\bibnamefont
  {Chung}},\ }\bibfield  {booktitle} {\emph {\bibinfo {booktitle}
  {{Proceedings, 15th Hellenic School and Workshops on Elementary Particle
  Physics and Gravity (CORFU2015): Corfu, Greece, September 1-25, 2015}}},\
  }\href {\doibase 10.3204/DESY-PROC-2015-02/woohyun_chung} {\bibfield
  {journal} {\bibinfo  {journal} {PoS}\ }\textbf {\bibinfo {volume}
  {CORFU2015}},\ \bibinfo {pages} {047} (\bibinfo {year} {2016})}\BibitemShut
  {NoStop}%
\bibitem [{\citenamefont {Brubaker}\ \emph {et~al.}(2017)\citenamefont
  {Brubaker} \emph {et~al.}}]{BenBs}%
  \BibitemOpen
  \bibfield  {author} {\bibinfo {author} {\bibfnamefont {B.~M.}\ \bibnamefont
  {Brubaker}} \emph {et~al.},\ }\href {\doibase 10.1103/PhysRevLett.118.061302}
  {\bibfield  {journal} {\bibinfo  {journal} {Phys. Rev. Lett.}\ }\textbf
  {\bibinfo {volume} {118}},\ \bibinfo {pages} {061302} (\bibinfo {year}
  {2017})},\ \Eprint {http://arxiv.org/abs/1610.02580} {arXiv:1610.02580
  [astro-ph.CO]} \BibitemShut {NoStop}%
\bibitem [{\citenamefont {Caldwel}\ \emph {et~al.}(2017)\citenamefont
  {Caldwel}, \citenamefont {Dvali}, \citenamefont {Majorovits}, \citenamefont
  {Millar}, \citenamefont {Raffelt}, \citenamefont {Redondo}, \citenamefont
  {Reimann}, \citenamefont {Simon},\ and\ \citenamefont {Steffen}}]{MADMAX}%
  \BibitemOpen
  \bibfield  {author} {\bibinfo {author} {\bibfnamefont {A.}~\bibnamefont
  {Caldwel}}, \bibinfo {author} {\bibfnamefont {G.}~\bibnamefont {Dvali}},
  \bibinfo {author} {\bibfnamefont {B.}~\bibnamefont {Majorovits}}, \bibinfo
  {author} {\bibfnamefont {A.}~\bibnamefont {Millar}}, \bibinfo {author}
  {\bibfnamefont {G.}~\bibnamefont {Raffelt}}, \bibinfo {author} {\bibfnamefont
  {J.}~\bibnamefont {Redondo}}, \bibinfo {author} {\bibfnamefont
  {O.}~\bibnamefont {Reimann}}, \bibinfo {author} {\bibfnamefont
  {F.}~\bibnamefont {Simon}}, \ and\ \bibinfo {author} {\bibfnamefont
  {F.}~\bibnamefont {Steffen}} (\bibinfo {collaboration} {MADMAX Working
  Group}),\ }\href {\doibase 10.1103/PhysRevLett.118.091801} {\bibfield
  {journal} {\bibinfo  {journal} {Phys. Rev. Lett.}\ }\textbf {\bibinfo
  {volume} {118}},\ \bibinfo {pages} {091801} (\bibinfo {year} {2017})},\
  \Eprint {http://arxiv.org/abs/1611.05865} {arXiv:1611.05865
  [physics.ins-det]} \BibitemShut {NoStop}%
\bibitem [{\citenamefont {Simanovskaia}\ and\ \citenamefont {van
  Bibber}(2015)}]{ADMXHF}%
  \BibitemOpen
  \bibfield  {author} {\bibinfo {author} {\bibfnamefont {M.}~\bibnamefont
  {Simanovskaia}}\ and\ \bibinfo {author} {\bibfnamefont {K.}~\bibnamefont {van
  Bibber}},\ }in\ \href {\doibase 10.3204/DESY-PROC-2015-02/simanovskaia_maria}
  {\emph {\bibinfo {booktitle} {{Proceedings, 11th Patras Workshop on Axions,
  WIMPs and WISPs (Axion-WIMP 2015): Zaragoza, Spain, June 22-26, 2015}}}}\
  (\bibinfo {year} {2015})\ pp.\ \bibinfo {pages} {157--163},\ \bibinfo {note}
  {[,157(2015)]}\BibitemShut {NoStop}%
\bibitem [{\citenamefont {Rybka}\ \emph {et~al.}(2015)\citenamefont {Rybka},
  \citenamefont {Wagner}, \citenamefont {Patel}, \citenamefont {Percival},
  \citenamefont {Ramos},\ and\ \citenamefont {Brill}}]{Rybka:2015aa}%
  \BibitemOpen
  \bibfield  {author} {\bibinfo {author} {\bibfnamefont {G.}~\bibnamefont
  {Rybka}}, \bibinfo {author} {\bibfnamefont {A.}~\bibnamefont {Wagner}},
  \bibinfo {author} {\bibfnamefont {K.}~\bibnamefont {Patel}}, \bibinfo
  {author} {\bibfnamefont {R.}~\bibnamefont {Percival}}, \bibinfo {author}
  {\bibfnamefont {K.}~\bibnamefont {Ramos}}, \ and\ \bibinfo {author}
  {\bibfnamefont {A.}~\bibnamefont {Brill}},\ }\href
  {https://link.aps.org/doi/10.1103/PhysRevD.91.011701} {\bibfield  {journal}
  {\bibinfo  {journal} {Physical Review D}\ }\textbf {\bibinfo {volume} {91}},\
  \bibinfo {pages} {011701} (\bibinfo {year} {2015})}\BibitemShut {NoStop}%
\bibitem [{\citenamefont {Barbieri}\ \emph {et~al.}(2017)\citenamefont
  {Barbieri}, \citenamefont {Braggio}, \citenamefont {Carugno}, \citenamefont
  {Gallo}, \citenamefont {Lombardi}, \citenamefont {Ortolan}, \citenamefont
  {Pengo}, \citenamefont {Ruoso},\ and\ \citenamefont
  {Speake}}]{Barbieri:2017aa}%
  \BibitemOpen
  \bibfield  {author} {\bibinfo {author} {\bibfnamefont {R.}~\bibnamefont
  {Barbieri}}, \bibinfo {author} {\bibfnamefont {C.}~\bibnamefont {Braggio}},
  \bibinfo {author} {\bibfnamefont {G.}~\bibnamefont {Carugno}}, \bibinfo
  {author} {\bibfnamefont {C.~S.}\ \bibnamefont {Gallo}}, \bibinfo {author}
  {\bibfnamefont {A.}~\bibnamefont {Lombardi}}, \bibinfo {author}
  {\bibfnamefont {A.}~\bibnamefont {Ortolan}}, \bibinfo {author} {\bibfnamefont
  {R.}~\bibnamefont {Pengo}}, \bibinfo {author} {\bibfnamefont
  {G.}~\bibnamefont {Ruoso}}, \ and\ \bibinfo {author} {\bibfnamefont {C.~C.}\
  \bibnamefont {Speake}},\ }\href {\doibase
  http://dx.doi.org/10.1016/j.dark.2017.01.003} {\bibfield  {journal} {\bibinfo
   {journal} {Physics of the Dark Universe}\ }\textbf {\bibinfo {volume}
  {15}},\ \bibinfo {pages} {135} (\bibinfo {year} {2017})}\BibitemShut
  {NoStop}%
\bibitem [{\citenamefont {Wilczek}(1987)}]{Wilczek:1987aa}%
  \BibitemOpen
  \bibfield  {author} {\bibinfo {author} {\bibfnamefont {F.}~\bibnamefont
  {Wilczek}},\ }\href {http://link.aps.org/doi/10.1103/PhysRevLett.58.1799}
  {\bibfield  {journal} {\bibinfo  {journal} {Physical Review Letters}\
  }\textbf {\bibinfo {volume} {58}},\ \bibinfo {pages} {1799} (\bibinfo {year}
  {1987})}\BibitemShut {NoStop}%
\bibitem [{\citenamefont {Sikivie}(1983)}]{Sikivie:1983aa}%
  \BibitemOpen
  \bibfield  {author} {\bibinfo {author} {\bibfnamefont {P.}~\bibnamefont
  {Sikivie}},\ }\href {http://link.aps.org/doi/10.1103/PhysRevLett.51.1415}
  {\bibfield  {journal} {\bibinfo  {journal} {Physical Review Letters}\
  }\textbf {\bibinfo {volume} {51}},\ \bibinfo {pages} {1415} (\bibinfo {year}
  {1983})}\BibitemShut {NoStop}%
\bibitem [{\citenamefont {Sikivie}(1985)}]{SecondSikivie}%
  \BibitemOpen
  \bibfield  {author} {\bibinfo {author} {\bibfnamefont {P.}~\bibnamefont
  {Sikivie}},\ }\href {\doibase 10.1103/PhysRevD.36.974,
  10.1103/PhysRevD.32.2988} {\bibfield  {journal} {\bibinfo  {journal} {Phys.
  Rev.}\ }\textbf {\bibinfo {volume} {D32}},\ \bibinfo {pages} {2988} (\bibinfo
  {year} {1985})},\ \bibinfo {note} {[Erratum: Phys.
  Rev.D36,974(1987)]}\BibitemShut {NoStop}%
\bibitem [{\citenamefont {Asztalos}\ \emph {et~al.}(2010)\citenamefont
  {Asztalos}, \citenamefont {Carosi}, \citenamefont {Hagmann}, \citenamefont
  {Kinion}, \citenamefont {van Bibber}, \citenamefont {Hotz}, \citenamefont
  {Rosenberg}, \citenamefont {Rybka}, \citenamefont {Hoskins}, \citenamefont
  {Hwang}, \citenamefont {Sikivie}, \citenamefont {Tanner}, \citenamefont
  {Bradley},\ and\ \citenamefont {Clarke}}]{Asztalos:2010aa}%
  \BibitemOpen
  \bibfield  {author} {\bibinfo {author} {\bibfnamefont {S.~J.}\ \bibnamefont
  {Asztalos}}, \bibinfo {author} {\bibfnamefont {G.}~\bibnamefont {Carosi}},
  \bibinfo {author} {\bibfnamefont {C.}~\bibnamefont {Hagmann}}, \bibinfo
  {author} {\bibfnamefont {D.}~\bibnamefont {Kinion}}, \bibinfo {author}
  {\bibfnamefont {K.}~\bibnamefont {van Bibber}}, \bibinfo {author}
  {\bibfnamefont {M.}~\bibnamefont {Hotz}}, \bibinfo {author} {\bibfnamefont
  {L.~J.}\ \bibnamefont {Rosenberg}}, \bibinfo {author} {\bibfnamefont
  {G.}~\bibnamefont {Rybka}}, \bibinfo {author} {\bibfnamefont
  {J.}~\bibnamefont {Hoskins}}, \bibinfo {author} {\bibfnamefont
  {J.}~\bibnamefont {Hwang}}, \bibinfo {author} {\bibfnamefont
  {P.}~\bibnamefont {Sikivie}}, \bibinfo {author} {\bibfnamefont {D.~B.}\
  \bibnamefont {Tanner}}, \bibinfo {author} {\bibfnamefont {R.}~\bibnamefont
  {Bradley}}, \ and\ \bibinfo {author} {\bibfnamefont {J.}~\bibnamefont
  {Clarke}},\ }\href {http://link.aps.org/doi/10.1103/PhysRevLett.104.041301}
  {\bibfield  {journal} {\bibinfo  {journal} {Physical Review Letters}\
  }\textbf {\bibinfo {volume} {104}},\ \bibinfo {pages} {041301} (\bibinfo
  {year} {2010})}\BibitemShut {NoStop}%
\bibitem [{\citenamefont {Rosenberg}(2015)}]{ADMX2}%
  \BibitemOpen
  \bibfield  {author} {\bibinfo {author} {\bibfnamefont {L.~J.}\ \bibnamefont
  {Rosenberg}},\ }in\ \href {\doibase 10.1073/pnas.1308788112} {\emph {\bibinfo
  {booktitle} {{Sackler Colloquium: Dark Matter Universe: On the Threshhold of
  Discovery Irvine, USA, October 18-20, 2012}}}}\ (\bibinfo {year}
  {2015})\BibitemShut {NoStop}%
\bibitem [{\citenamefont {McAllister}\ \emph
  {et~al.}(2016{\natexlab{c}})\citenamefont {McAllister}, \citenamefont
  {Parker},\ and\ \citenamefont {Tobar}}]{McAllister:2016aa}%
  \BibitemOpen
  \bibfield  {author} {\bibinfo {author} {\bibfnamefont {B.~T.}\ \bibnamefont
  {McAllister}}, \bibinfo {author} {\bibfnamefont {S.~R.}\ \bibnamefont
  {Parker}}, \ and\ \bibinfo {author} {\bibfnamefont {M.~E.}\ \bibnamefont
  {Tobar}},\ }\href {http://link.aps.org/doi/10.1103/PhysRevLett.117.159901}
  {\bibfield  {journal} {\bibinfo  {journal} {Physical Review Letters}\
  }\textbf {\bibinfo {volume} {117}},\ \bibinfo {pages} {159901} (\bibinfo
  {year} {2016}{\natexlab{c}})}\BibitemShut {NoStop}%
\bibitem [{\citenamefont {Lievens}\ \emph {et~al.}(2013)\citenamefont
  {Lievens}, \citenamefont {Stoilova},\ and\ \citenamefont {der
  Jeugt}}]{Lievens:2013aa}%
  \BibitemOpen
  \bibfield  {author} {\bibinfo {author} {\bibfnamefont {S.}~\bibnamefont
  {Lievens}}, \bibinfo {author} {\bibfnamefont {N.}~\bibnamefont {Stoilova}}, \
  and\ \bibinfo {author} {\bibfnamefont {J.~V.}\ \bibnamefont {der Jeugt}},\
  }\href {http://dx.doi.org/10.1016/j.physletb.2013.01.013} {\bibfield
  {journal} {\bibinfo  {journal} {Physics Letters B}\ } (\bibinfo {year}
  {2013})}\BibitemShut {NoStop}%
\bibitem [{\citenamefont {McAllister}\ \emph {et~al.}(2017)\citenamefont
  {McAllister}, \citenamefont {Parker}, \citenamefont {Ivanov},\ and\
  \citenamefont {Tobar}}]{XSWISP}%
  \BibitemOpen
  \bibfield  {author} {\bibinfo {author} {\bibfnamefont {B.~T.}\ \bibnamefont
  {McAllister}}, \bibinfo {author} {\bibfnamefont {S.~R.}\ \bibnamefont
  {Parker}}, \bibinfo {author} {\bibfnamefont {E.~N.}\ \bibnamefont {Ivanov}},
  \ and\ \bibinfo {author} {\bibfnamefont {M.~E.}\ \bibnamefont {Tobar}},\
  }\href@noop {} {\  (\bibinfo {year} {2017})},\ \Eprint
  {http://arxiv.org/abs/1510.05775} {arXiv:1510.05775 [physics.ins-det]}
  \BibitemShut {NoStop}%
\bibitem [{\citenamefont {Hansen}(1938)}]{reen0}%
  \BibitemOpen
  \bibfield  {author} {\bibinfo {author} {\bibfnamefont {W.}~\bibnamefont
  {Hansen}},\ }\href@noop {} {\bibfield  {journal} {\bibinfo  {journal}
  {Journal of Applied Physics}\ }\textbf {\bibinfo {volume} {9}},\ \bibinfo
  {pages} {654} (\bibinfo {year} {1938})}\BibitemShut {NoStop}%
\bibitem [{\citenamefont {Floch}\ \emph {et~al.}(2013)\citenamefont {Floch},
  \citenamefont {Fan}, \citenamefont {Aubourg}, \citenamefont {Cros},
  \citenamefont {Carvalho}, \citenamefont {Shan}, \citenamefont {Bourhill},
  \citenamefont {Ivanov}, \citenamefont {Humbert}, \citenamefont {Madrangeas},\
  and\ \citenamefont {Tobar}}]{reen1}%
  \BibitemOpen
  \bibfield  {author} {\bibinfo {author} {\bibfnamefont {J.-M.~L.}\
  \bibnamefont {Floch}}, \bibinfo {author} {\bibfnamefont {Y.}~\bibnamefont
  {Fan}}, \bibinfo {author} {\bibfnamefont {M.}~\bibnamefont {Aubourg}},
  \bibinfo {author} {\bibfnamefont {D.}~\bibnamefont {Cros}}, \bibinfo {author}
  {\bibfnamefont {N.}~\bibnamefont {Carvalho}}, \bibinfo {author}
  {\bibfnamefont {Q.}~\bibnamefont {Shan}}, \bibinfo {author} {\bibfnamefont
  {J.}~\bibnamefont {Bourhill}}, \bibinfo {author} {\bibfnamefont
  {E.}~\bibnamefont {Ivanov}}, \bibinfo {author} {\bibfnamefont
  {G.}~\bibnamefont {Humbert}}, \bibinfo {author} {\bibfnamefont
  {V.}~\bibnamefont {Madrangeas}}, \ and\ \bibinfo {author} {\bibfnamefont
  {M.}~\bibnamefont {Tobar}},\ }\href@noop {} {\bibfield  {journal} {\bibinfo
  {journal} {Review of Scientific Instruments}\ }\textbf {\bibinfo {volume}
  {84}},\ \bibinfo {pages} {125114} (\bibinfo {year} {2013})}\BibitemShut
  {NoStop}%
\bibitem [{\citenamefont {Fujisawa}(1958)}]{reen2}%
  \BibitemOpen
  \bibfield  {author} {\bibinfo {author} {\bibfnamefont {K.}~\bibnamefont
  {Fujisawa}},\ }\href@noop {} {\bibfield  {journal} {\bibinfo  {journal} {IRE
  Trans. On Microwave Theory and Techniques}\ }\textbf {\bibinfo {volume}
  {6}},\ \bibinfo {pages} {344} (\bibinfo {year} {1958})}\BibitemShut {NoStop}%
\bibitem [{\citenamefont {Goryachev}\ and\ \citenamefont
  {Tobar}(2015{\natexlab{a}})}]{Goryachev:2015aa}%
  \BibitemOpen
  \bibfield  {author} {\bibinfo {author} {\bibfnamefont {M.}~\bibnamefont
  {Goryachev}}\ and\ \bibinfo {author} {\bibfnamefont {M.~E.}\ \bibnamefont
  {Tobar}},\ }\href {http://stacks.iop.org/1367-2630/17/i=2/a=023003}
  {\bibfield  {journal} {\bibinfo  {journal} {New Journal of Physics}\ }\textbf
  {\bibinfo {volume} {17}},\ \bibinfo {pages} {023003} (\bibinfo {year}
  {2015}{\natexlab{a}})}\BibitemShut {NoStop}%
\bibitem [{\citenamefont {Goryachev}\ and\ \citenamefont
  {Tobar}(2015{\natexlab{b}})}]{Goryachev:2015ab}%
  \BibitemOpen
  \bibfield  {author} {\bibinfo {author} {\bibfnamefont {M.}~\bibnamefont
  {Goryachev}}\ and\ \bibinfo {author} {\bibfnamefont {M.~E.}\ \bibnamefont
  {Tobar}},\ }\href
  {http://scitation.aip.org/content/aip/journal/jap/118/20/10.1063/1.4936268}
  {\bibfield  {journal} {\bibinfo  {journal} {Journal of Applied Physics}\
  }\textbf {\bibinfo {volume} {118}},\ \bibinfo {pages} {204504} (\bibinfo
  {year} {2015}{\natexlab{b}})}\BibitemShut {NoStop}%
\bibitem [{\citenamefont {Goryachev}\ and\ \citenamefont
  {Tobar}(2016)}]{Goryachev:2016aa}%
  \BibitemOpen
  \bibfield  {author} {\bibinfo {author} {\bibfnamefont {M.}~\bibnamefont
  {Goryachev}}\ and\ \bibinfo {author} {\bibfnamefont {M.~E.}\ \bibnamefont
  {Tobar}},\ }\href {http://link.aps.org/doi/10.1103/PhysRevApplied.6.064006}
  {\bibfield  {journal} {\bibinfo  {journal} {Physical Review Applied}\
  }\textbf {\bibinfo {volume} {6}},\ \bibinfo {pages} {064006} (\bibinfo {year}
  {2016})}\BibitemShut {NoStop}%
\bibitem [{\citenamefont {Bassan}\ \emph {et~al.}(2008)\citenamefont {Bassan},
  \citenamefont {Ballantini}, \citenamefont {Chincarini}, \citenamefont
  {Gemme}, \citenamefont {Iannuzzi}, \citenamefont {Moleti}, \citenamefont
  {Parodi},\ and\ \citenamefont {Vaccarone}}]{Bassan2008}%
  \BibitemOpen
  \bibfield  {author} {\bibinfo {author} {\bibfnamefont {M.}~\bibnamefont
  {Bassan}}, \bibinfo {author} {\bibfnamefont {R.}~\bibnamefont {Ballantini}},
  \bibinfo {author} {\bibfnamefont {A.}~\bibnamefont {Chincarini}}, \bibinfo
  {author} {\bibfnamefont {G.}~\bibnamefont {Gemme}}, \bibinfo {author}
  {\bibfnamefont {M.}~\bibnamefont {Iannuzzi}}, \bibinfo {author}
  {\bibfnamefont {A.}~\bibnamefont {Moleti}}, \bibinfo {author} {\bibfnamefont
  {R.~F.}\ \bibnamefont {Parodi}}, \ and\ \bibinfo {author} {\bibfnamefont
  {R.}~\bibnamefont {Vaccarone}},\ }\href@noop {} {\bibfield  {journal}
  {\bibinfo  {journal} {Journal of Physics: Conference Series}\ }\textbf
  {\bibinfo {volume} {122}},\ \bibinfo {pages} {012031} (\bibinfo {year}
  {2008})}\BibitemShut {NoStop}%
\bibitem [{\citenamefont {McAllister}\ \emph
  {et~al.}(2016{\natexlab{d}})\citenamefont {McAllister}, \citenamefont
  {Parker},\ and\ \citenamefont {Tobar}}]{RCAxion}%
  \BibitemOpen
  \bibfield  {author} {\bibinfo {author} {\bibfnamefont {B.~T.}\ \bibnamefont
  {McAllister}}, \bibinfo {author} {\bibfnamefont {S.~R.}\ \bibnamefont
  {Parker}}, \ and\ \bibinfo {author} {\bibfnamefont {M.~E.}\ \bibnamefont
  {Tobar}},\ }\href {\doibase 10.1103/PhysRevD.94.042001} {\bibfield  {journal}
  {\bibinfo  {journal} {Phys. Rev. D}\ }\textbf {\bibinfo {volume} {94}},\
  \bibinfo {pages} {042001} (\bibinfo {year} {2016}{\natexlab{d}})}\BibitemShut
  {NoStop}%
\end{thebibliography}
%

\end{document}